\documentclass[aps,prl,twocolumn,showpacs,superscriptaddress,groupedaddress]{revtex4}  

\usepackage{graphicx}  
\usepackage{dcolumn}   
\usepackage{bm}        
\usepackage{amssymb}   
\usepackage[utf8]{inputenc}
\usepackage{gensymb}
\newcommand{\mic}{\,\mu\mathrm{m}}
\newcommand{\nm}{\,\mathrm{nm}}

\newcommand{\kev}{\,\mathrm{keV}}
\newcommand{\fs}{\,\mathrm{fs}}

\newcommand{\pc}{\,\mathrm{pC}}

\begin{document}



\title{Anticorrelated emission of high-harmonics and fast electron beams from plasma mirrors}
\author{Maïmouna Bocoum,$^{1,*}$ Maxence Thévenet,$^{1}$ Frederik Böhle,$^{1}$ Benoît Beaurepaire,$^{1}$ Aline Vernier,$^{1}$ Aurélie Jullien,$^{1}$ Jérôme Faure,$^{1}$ Rodrigo Lopez-Martens}

\address{
Laboratoire d'Optique Appliquée, ENSTA ParisTech, CNRS, Ecole polytechnique, Université Paris-Saclay, 828 bd des Maréchaux, 91762 Palaiseau cedex France
\\
$^*$Corresponding author: maimouna.bocoum@ensta-paristech.fr
}
\date{\today}

\begin{abstract}
We report for the first time on the anticorrelated emission of high-order harmonics and energetic electron beams from a solid-density plasma with a sharp vacuum interface$-$plasma mirror$-$driven by an intense ultrashort laser pulse. We highlight the key role played by the nanoscale structure of the plasma surface during the interaction by measuring the spatial and spectral properties of harmonics and electron beams emitted by a plasma mirror. We show that the nanoscale behavior of the plasma mirror can be controlled by tuning the scale length of the electronic density gradient, which is measured in-situ using spatial-domain interferometry.  

\end{abstract}

\maketitle



Over the past 30 years, solid-density plasmas driven by intense femtosecond (fs) pulses, so-called plasma mirrors, have been successfully tested as a source of high-order harmonics and attosecond XUV pulses in a number of experiments~\cite{burnett1977harmonic,carman1981observation,von1995generation,tarasevitch2000generation,quere2006coherent,nomura2009attosecond,easter2010high,borot2011high,heissler2012few,wheeler2012attosecond}, where the laser intensity typically exceeds a few $10^{14}\,\mathrm{W/cm}^2$. Other experiments have shown it is also possible to accelerate energetic electrons from plasma mirrors for intensities above $10^{16}\,\mathrm{W/cm}^2$~\cite{bastiani1997experimental,brandl2009directed,mordovanakis2009quasimonoenergetic}. Attempting to understand each of these experimental observations invariably points to the key role played by the plasma-vacuum interface during the interaction both on the nanoscale spatially and on the sub-laser-cycle scale temporally~\cite{thaury2007plasma,borot2012attosecond}.

\begin{figure}
\includegraphics[scale=0.12]{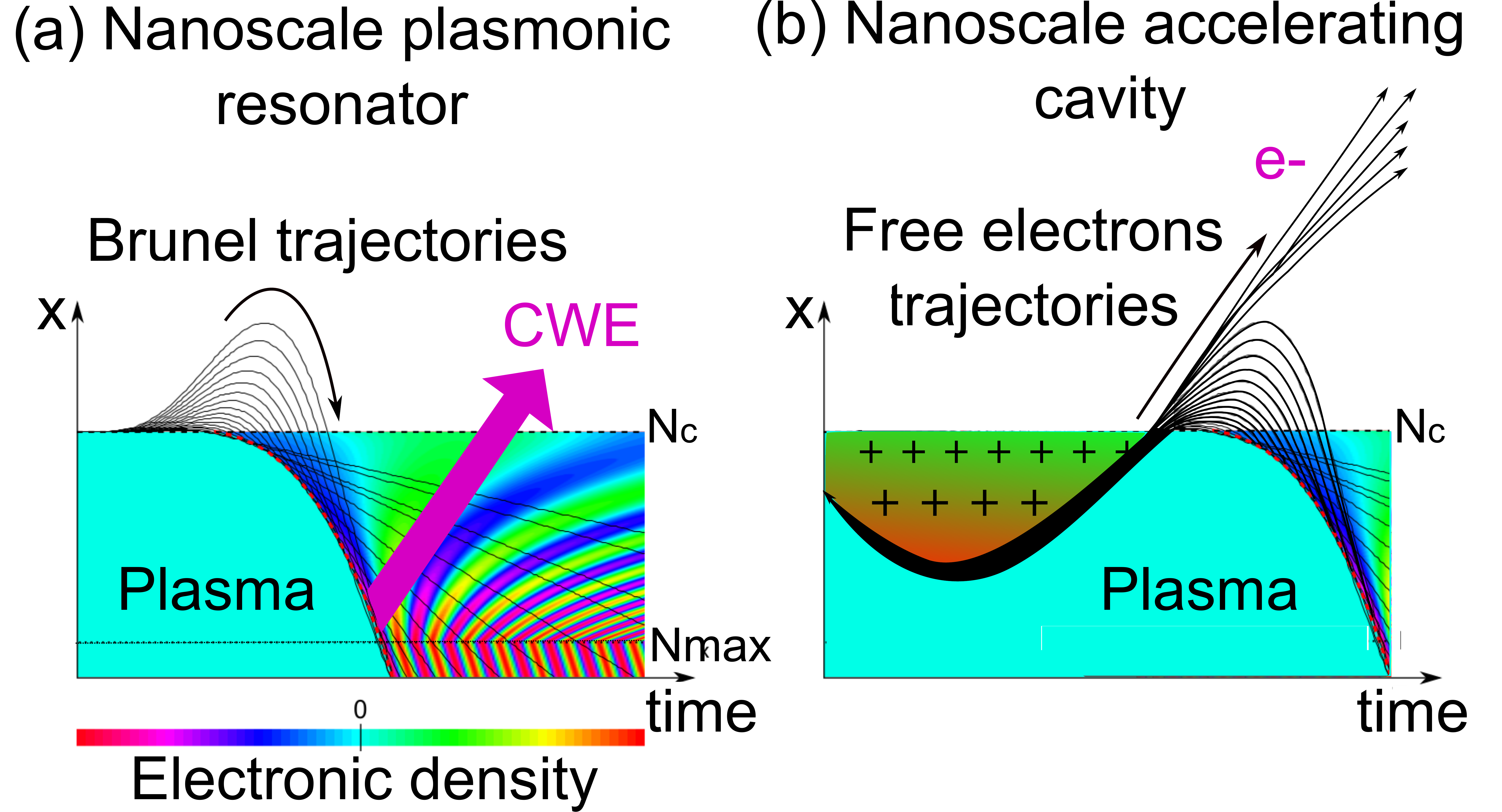}
\caption{\label{fig:schemaNanoStructures}Diagrams of nanoscale plasma mirror surface structures leading to (a) CWE: electrons are pulled toward the vacuum and are sent back to the plasma where they excite high-frequency plasma waves, which radiate high-order harmonics (b) electron acceleration on the sub-laser-cycle time scale: electrons are accelerated in the density gradient and escape from the plasma.}
\end{figure}

It is commonly assumed that the electronic density at the plasma mirror surface decreases exponentially from solid to vacuum over a distance $L_g$, also called density gradient. When the laser pulse reflects on this plasma mirror, for every oscillation of the laser field, some electrons are driven towards vacuum and sent back to the plasma~\cite{lichters1996short,liseykina2015collisionless}. These bunches of so-called Brunel electrons~\cite{brunel1987not} impulsively excite collective high-frequency plasma oscillations in the density gradient that lead to the emission of XUV radiation through linear mode conversion~\cite{thaury2010high}. As illustrated in Fig~\ref{fig:schemaNanoStructures}(a), each position $x$ of the plasma behaves as a nanoscale oscillator of frequency $\omega_p(x) = \omega_0\sqrt{n_e(x)/n_{c}}$ where $\omega_0$ is the driving laser angular frequency, $n_e$ the local electronic density at position $x$ and $n_{c}$ the critical density. This periodic mechanism, called Coherent Wake Emission (CWE), leads to efficient high harmonics generation for very short plasma scale lengths, typically $L_g\sim \lambda/100$~\cite{thaury2010high}, even for sub-relativistic intensities, $a_0<1$, where $a_0=eA_0/mc$ is the normalized vector potential, $e$ and $m$ the electron charge and mass and $c$ the speed of light. However, the efficiency significantly drops for $L_g >> \lambda/20$~\cite{tarasevitch2000generation,quere2006coherent,kahaly2013direct,thaury2010high}. At higher intensities $a_0\gg1$, the Relativistic Oscillating Mirror (ROM) becomes the dominant mechanism for harmonic generation~\cite{lichters1996short,bulanov1994interaction}. 

A fraction of electrons do not follow Brunel-like trajectories: they are accelerated in the density gradient towards vacuum and escape the plasma, as illustrated in Fig~\ref{fig:schemaNanoStructures}(b). Depending on the interaction conditions, the final energy and angular spread of these electrons can be influenced by plasma waves below the critical surface~\cite{bastiani1997experimental}, interference fields created by the incident and reflected laser beams \cite{tian2012electron,mordovanakis2009quasimonoenergetic,naumova2004attosecond}, betatron-like motion at the plasma surface~\cite{chen2006surface} or even direct laser acceleration in vacuum \cite{thevenet2015}. Here again, the plasma scale length plays a critical role: enhanced electron generation is observed typically for $0.1<L_g/\lambda <1$\cite{bastiani1997experimental,geindre2010electron,mordovanakis2009quasimonoenergetic} or sometimes even for $L_g/\lambda >1$~\cite{wang2010angular,cai2004double,yu2000electron}. 
To our knowledge, the anticorrelated emission of harmonics and fast electrons from plasma mirrors has never been investigated experimentally. In this letter, through a controlled pump-probe experiment using sub-relativistic femtosecond laser pulses, we directly observe the transition from a confined plasma that can efficiently emit laser harmonics to an extended plasma structure that accelerates fast electrons into vacuum up to a few hundreds $\mathrm{keV}$ energies, where the laser interference field only plays a second role.

The experiment was carried out using the ``Salle Noire'' laser system at the Laboratory of Applied Optics (LOA) delivering up to 3\,mJ energy, 30\,fs pulses at 1kHz repetition rate with high temporal contrast ($> 10^{-10}$)~\cite{jullien2014carrier}. The $p$-polarized pulses are focused  down to $1.7\mic$ FWHM spot size onto an optically flat fused silica target ($\sim 250 n_c$), leading to peak intensities on target $\simeq 10^{18}\,\mathrm{W/cm}^2$ ($a_0\simeq0.7$) for an incidence angle $\theta_{L} = 49.3^{\circ}$, with high repeatability at 1~kHz~\cite{borot2014high}. $5\%$ of the main beam is picked off and focused down to 5 times the main beam spot size on target in order to induce homogeneous plasma expansion at the surface (see also Supp. Mat \cite{Suppmat}). The plasma scale length $L_g$ can then be varied by changing the relative delay between this prepulse and the main high-intensity pulse. We use Spatial Domain Interferometry (SDI)~\cite{bocoum2015spatial} to estimate the plasma expansion velocity $c_s$ and find $c_s = dL_g/dt= 10.8\pm 1.1,\mathrm{nm/ps}$ for a prepulse intensity of $\simeq 3.5 \times 10^{14}\,\mathrm{W/cm}^2$ ($a_0\simeq 0.013$). 

Harmonics emitted in the specular direction are sent into a home-made XUV spectrometer where the harmonic spectrum is resolved in the horizontal plane and the harmonic beam divergence in the vertical direction using a coupled MCP and phosphor screen detector. At the same time, a $6\times 17\,\mathrm{cm}$ Lanex screen was positioned $10\,\mathrm{cm}$ away parallel to the target surface without blocking the specular direction.  The angular electron emission profile in this geometry was recorded as a function of $\theta \in [-20^{\circ} \ 30^{\circ}]$, the angle with respect to target normal in the plane of incidence and $\phi\in [-20^{\circ} \ 20^{\circ}]$, the angle with respect to target normal in the tangential plane. Note that the Lanex screen only detects electrons with energies larger than 150~keV~\cite{glinec2006absolute}. The Lanex screen could also be replaced by an electron spectrometer for characterizing the electron energy distribution.
\begin{figure}[!h]
\includegraphics[scale=0.4]{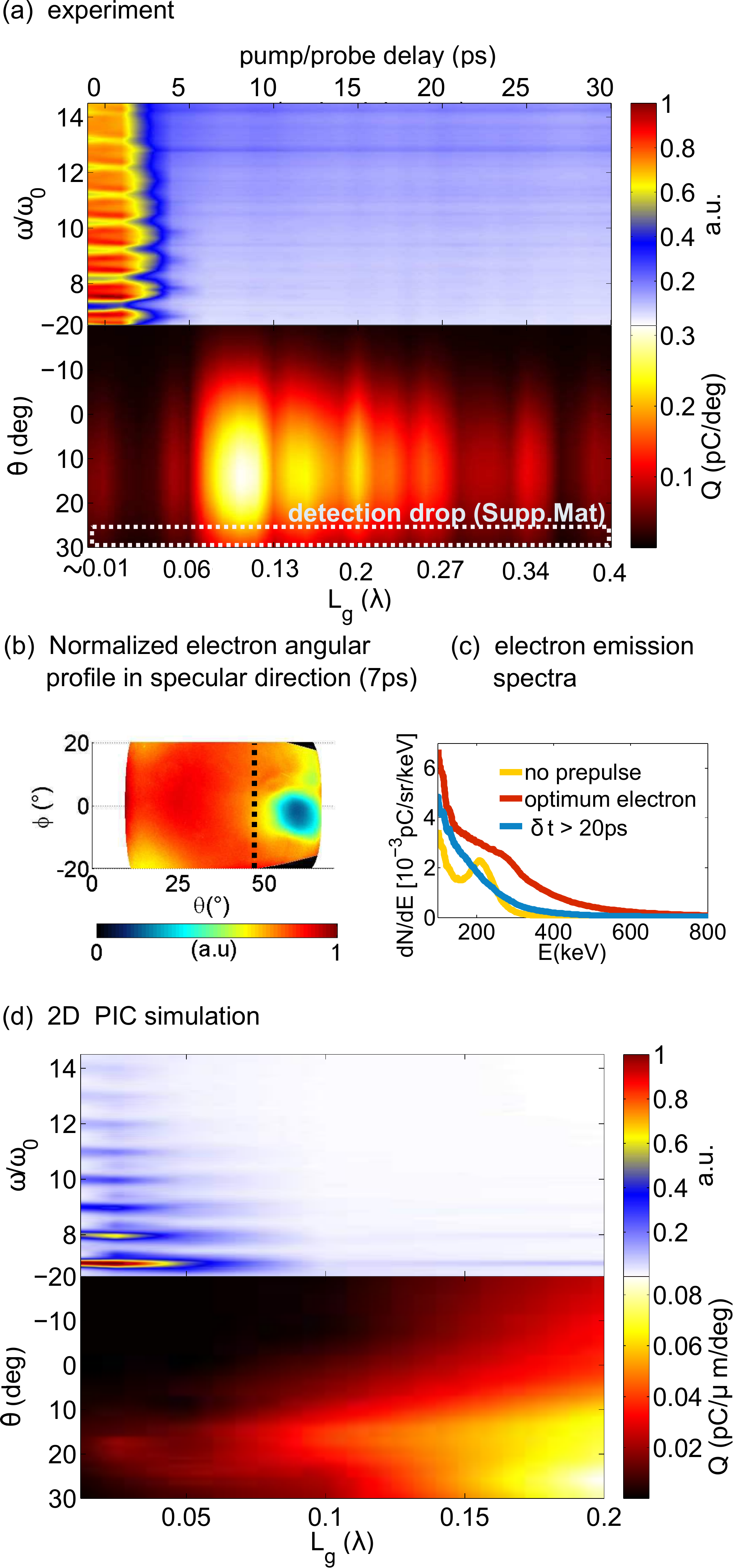}
\caption{\label{fig:anti-corr}(a) Experimental harmonic spectra and electron angular emission profiles as a function of pump-probe delay (top axis) between the prepulse and the main pulse. The electron signal was integrated along the tangential coordinate $\phi$. The corresponding plasma scale length $L_g$ (bottom axis) was extracted from the plasma expansion velocity $c_s = 10.8\,\mathrm{nm/ps}$ measured by SDI~\cite{bocoum2015spatial}; (b) Electron angular distribution when the Lanex is placed perpendicular to the specular direction and after deconvolution (\textit{Supp Mat}) (c) Electron energy spectra for three typical delays; (d) Same as (a) for 2D PIC simulations with $a_0=0.4$, and gradient length $L_g\in[0.01\lambda;0.2\lambda]$. }
\end{figure}

Figure~\ref{fig:anti-corr}(a) shows the harmonic spectrum and the electron signal as a function of pump-probe delay, hence the gradient length. The harmonic signal was integrated along the divergence angle. The plasma scale length calculated from the plasma expansion velocity is indicated on the bottom axis. The first striking result is that harmonics are generated efficiently for pump-probe delays below 4\,ps, corresponding to $L_g \le 0.05\lambda$. The spectrum extends up to the plasma frequency cut-off $\omega_c/\omega_0 = 16$ and its divergence is about $1/10^{th}$ that of the driving laser beam, which is the typical signature of CWE~\cite{kahaly2013direct}. The plasma frequency cut-off confirms that Brunel electrons can efficiently excite collective plasma oscillations and therefore that the initial plasma scale length should be on the order of $L_g\sim 0.01\lambda$ \cite{thaury2010high} rather then rigorously $0\lambda$. This also indicates that the temporal contrast close to the pulse peak does not allow us to explore arbitrarily small plasma scale lengths. The drop in CWE emission efficiency with increasing density gradient has already been observed experimentally and is theoretically predicted to be in the range $0.02 < L_g/\lambda <0.1$~\cite{kahaly2013direct,tarasevitch2007transition} depending on laser intensity~\cite{thaury2010high}. This can be explained with 1D considerations: the minimum time required to excite plasma waves from the critical surface $x = x_c$ to the location of maximum density $x = x_{max}$ is $\Delta t = (L_g/c)\log(n_{max}/n_c)$, which should be less than the laser period in order to prevent cycle-to-cycle destructive interferences. For traveling electrons, this limit reads $L_g \le 0.17\lambda$. In our case, the drop in efficiency occurs at much lower values around $L_g\sim 0.05\lambda$ because the electronic perturbation propagates at less than $c$ and the initial perturbation strength (i.e. amplitude of plasma waves) decreases with $L_g$~\cite{thaury2010high}.
The second striking result is that a maximum electron signal is reached for a delay of 8\,ps ($L_g \sim 0.1 \lambda$), where harmonic emission is negligible. The ejected electrons form a large spot between $10^{\circ}$ and $20^{\circ}$ and drop at the edge of the Lanex at $\sim 30^{\circ}$. This drop in signal is a geometrical artifact due to the anisotropic emission of the Lanex screen~\cite{glinec2006absolute} (see Supp. Mat. for details). Fig~\ref{fig:anti-corr}(b) shows the full electron angular distribution for a delay of $\sim7\,\mathrm{ps}$, obtained by moving the Lanex screen perpendicular to the specular direction. The distribution displays a hole close to the specular direction, presumably formed by the ponderomotive force of the reflected laser pulse \cite{thevenet2015,quesnel1998theory,mordovanakis2009quasimonoenergetic,tian2012electron}. Using the Lanex calibration~\cite{glinec2006absolute}, we estimate that the ejected charge reaches a maximum of $\sim 11\,\mathrm{pC}$ compared to $\sim 2\,\mathrm{pC}$ at zero delay.  Fig~\ref{fig:anti-corr}(c) shows electron spectra respectively without prepulse, for the optimal delay for electronic emission, and after $20\,\mathrm{ps}$. Hence, electrons can be effectively accelerated up to $\sim 600\,\mathrm{keV}$ at the optimal density gradient.

\begin{figure}[ht!]
\includegraphics[scale=0.7]{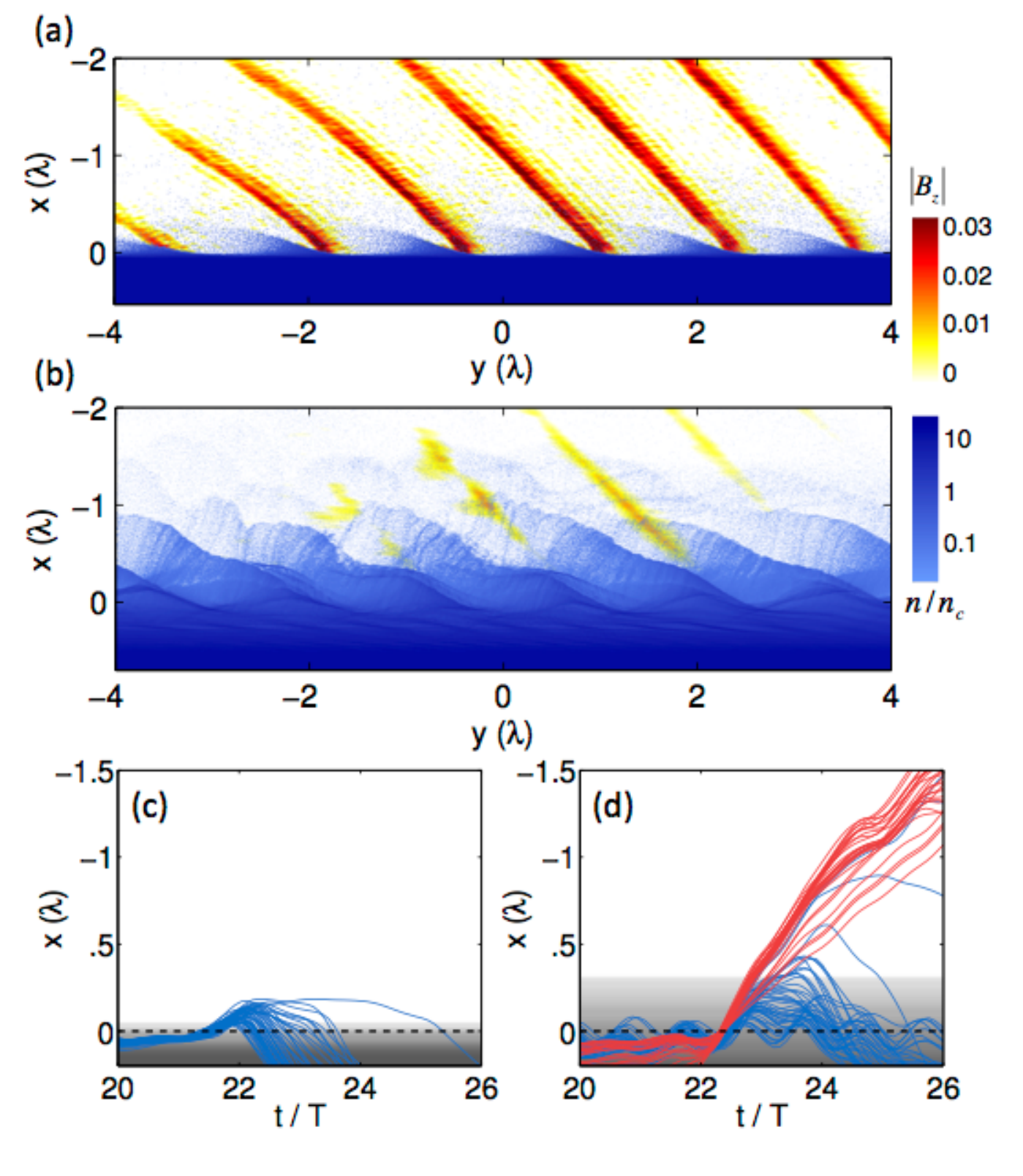}
\caption{(a) Snapshot from the 2D PIC simulation for $L_g=\lambda/40$. Blue: electron density (log-scale). Yellow-red: reflected harmonic field (a Fourier filter was applied to keep only harmonic orders $\geq 5\omega_0$). The harmonic field comes out as a train of attosecond pulses; (b) Same as (a) for $L_g=\lambda/5$ (same instant, same colour scale); (c) and (d) typical electron trajectories for $L_g=\lambda/40$ and $L_g=\lambda/5$ respectively. $x$ is the coordinate normal to the plasma. The grey scale stands for the plasma initial density and the black dotted line ($x=0$) shows the position of the critical density. The electrons represented here interact with the laser around its maximum ($t=22T$). Red trajectories stand for ejected electrons.}
\label{fig:density}
\end{figure}

To summarize, we observe that the emission of harmonics and electrons is anti-correlated when changing the gradient scale length. These experimental results were confronted to 2D Particle-In-Cell simulations, in which a $\lambda=800\nm$, $30 \fs$ pulse is focused onto an overdense plasma ($n_{max}=250n_c$) with immobile ions. The plasma density decreases exponentially with various scale lengths, from $L_g = 0.01\lambda$ to $0.2 \lambda$. The plasma density is cut at $n_b=n_c/5$, so that the plasma boundary is defined by $x_b=- \log 5 L_g$. The laser amplitude is $a_0=0.4$ and the incidence angle $45^\circ$. A good spatial resolution is required for simulating CWE harmonics, so we use $\delta x = \lambda/420$. In the simulations, electrons are detected at $9\lambda$ away from the critical surface and only electrons with energies $>150$~keV are detected (as in the experiment). As illustrated in Fig~\ref{fig:anti-corr}(d), the PIC simulations qualitatively reproduce our experimental observations: the CWE emission efficiency decreases for $L_g> 0.05\lambda $ and the effective ejected electron charge increases up to $\sim 3\,\pc.\mic^{-1}$ for $L_g = 0.2\lambda$ compared to $0.12\,\pc.\mic^{-1}$ when $L_g = 0.01\lambda$ (\textit{i.e.} $\simeq 10\pc$ and $0.7\pc$ respectively, for a $3.4\mic$ spot size FWHM).
The electron angular distribution was plotted over the range $\theta \in [-20^{\circ} \ 30^{\circ}]$ for a direct comparison with experiment. Here again, there is a very good agreement with the experiment, with a large divergence $10\pc$ beam ejected at $\sim 30^{\circ}$ when $L_g \sim 0.2\lambda$. Note that PIC simulations were first performed with the experimental vacuum laser amplitude $a_0=0.8$, but a strong harmonic emission attributed to the ROM emission mechanism~\cite{kahaly2013direct} persisted for longer gradients. These simulations at high intensities suggested a correlation between ROM harmonics and electron ejection, as opposed to the anti-correlation that we observed. In our experiment, ROM emission does not occur and the harmonics are due to CWE. This indicates that the laser intensity at focus is not high enough to support ROM emission \cite{thaury2007plasma}. Therefore, in the simulations, the beam spot size was doubled without changing the pulse energy, i.e. $a_0$ was decreased to 0.4, to reproduce the anticorrelated behavior. Note that our overestimation of the experimental intensity on target may be due to a slight defocusing of the laser on target or debris reducing the overall transmission of the focusing optic, a standard problem with high repetition rate laser-plasma interaction experiments using tight focusing.

Figure~\ref{fig:density} shows a comparison of 2D PIC simulations with a gradient length optimized for harmonic emission ($L_g=\lambda/40$) and electron emission ($L_g=\lambda/5$), respectively. In panels (a) and (b), one can clearly see oscillations of the electron surface at the laser period. Strong harmonic generation can be seen in panel (a). The corresponding electron trajectories are shown in panel (c), where the $x$ coordinate (normal to the target) of electrons is plotted along time. For clarity, a single bunch of electrons is represented here, that interacts with the laser around its temporal maximum ($t=22T$, where $T$ is the optical period) in the center of the interference pattern. One can clearly see Brunel-like trajectories: electrons make a short excursion in vacuum before being driven back to the plasma where they trigger plasma waves. In panel (b), the amplitude of these oscillations is greater and layers of electrons are ejected from the plasma surface. The corresponding electron trajectories are plotted in panel (d). Once again, a bunch of electrons was selected for clarity. A fraction of these electrons (in red) escape from the plasma and propagate into vacuum in the interference pattern with a velocity $\simeq c/2$.



\begin{figure}[ht!]
\includegraphics[scale=0.6]{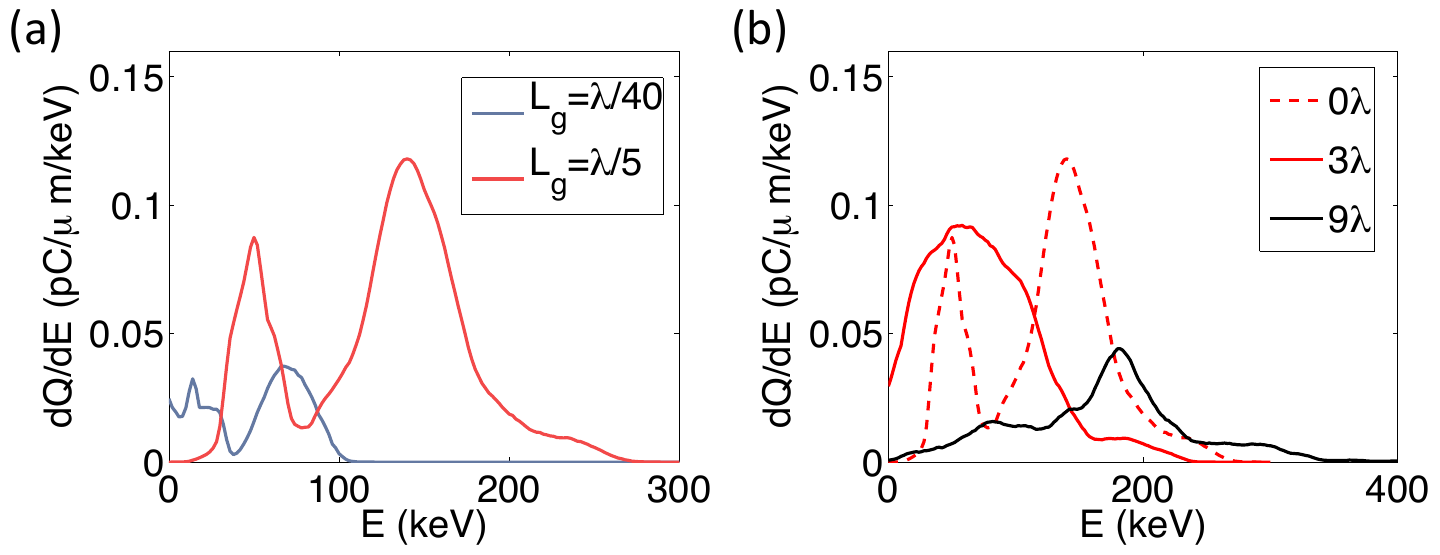}
\caption{\label{fig:spectrasHHG}{Simulated ejected electron spectra at the plasma boundary $x_b$ for $L_g=\lambda/40$ (grey) and $L_g=\lambda/5$ (red) as they cross the plasma critical surface; (b) Ejected electron spectra for $L_g=\lambda/5$ as they cross the plasma critical surface (dotted line), $3\lambda$ and $9 \lambda$ away from the plasma surface (respectively red and black solid line).}}
\end{figure}

For each laser cycle, the ejection mechanism can be described as follows: (i) the laser electric field pushes electrons inside the plasma, while the heavy ions stay in place, creating a charge separation electrostatic field, i.e. a  plasma capacitor which can give potential energy to electrons. (ii) Half a cycle later, the laser field changes sign and both the capacitor and the laser electric force pull and accelerate electrons towards vacuum. Assuming that all the electrons originating from $x<0$ (where $n=n_c$) are pushed towards $x\geq 0$, the electrostatic potential of the remaining ions can be calculated using Poisson's equation $\Delta V_P = -n_c/\epsilon_0e^{x/L_g}$, and reads $V_P = -n_cL_g^2/\epsilon_0$. Therefore, electrons are expected to gain more energy from the plasma for longer gradients. Fig.~\ref{fig:spectrasHHG}(a) shows the spectrum of ejected electrons when they cross the plasma boundary at $x_b$. The average energy is much higher for longer gradients, thus confirming our predictions. Hence, the plasma serves as an injector of electrons into the reflecting laser~\cite{thevenet2015}.
In order to determine whether the electrons are mainly accelerated in the plasma or in the interference pattern, we plot the simulated electron spectra at the plasma border, at $3.3\lambda$ and $9\lambda$ away from the plasma in Fig.~\ref{fig:spectrasHHG}(b). Within this range, no net energy gain can be observed from the electromagnetic wave in vacuum, we conclude that the energy gain is mostly due to acceleration inside the plasma gradient. However, further away from the plasma at $9\lambda$, the electron spectrum broadens and the tail of the distribution reaches $400\kev$, which could be the signature of ponderomotive~\cite{mordovanakis2009quasimonoenergetic,naumova2004attosecond} and/or stochastic heating in the interference pattern~\cite{shen02}. The formation of a hole in the experimental electron angular emission profile (see Fig~\ref{fig:anti-corr}(b)) and the absence of a beaming as seen in \cite{thevenet2015} is more evidence that the interaction between the accelerated electrons and the laser is purely ponderomotive. Finally, from simulations and experiments, we also conclude that for $a_0 <1$ and $L_g \sim 0.1\lambda$, electrons cannot be accelerated by plasma waves related to the CWE mechanism, as suggested in \cite{bastiani1997experimental}, otherwise electron and harmonic emission would be optimal simultaneously.

To conclude, we observe for the first time the transition from high-harmonic emission to fast electron ejection as the electronic density gradient increases at the surface of a plasma mirror driven at sub-relativistic laser intensity. Our measurements reveal that both processes cannot occur simultaneously for the same density gradient. For sharp gradients ($L_g < 0.05\lambda $), electrons drive oscillations in a confined plasma, leading to efficient coherent harmonic emission in their wake. For softer gradients, electrons can be efficiently accelerated out of the plasma by the space-charge field created for $L_g\sim 0.1 \lambda$. Although the interaction with the reflected laser field thermalizes the electron population and reshapes the spatial emission profile via ponderomotive interactions, most of the acceleration occurs inside the plasma density gradient. As the gradient length increases by $\sim 40\,\mathrm{nm}$, the plasma mirror behavior switches from a collection of efficient XUV resonators to a nanoscale electron accelerator. 

We thank N. Naumova for useful discussions.
This work was funded by the Agence Nationale pour la Recherche (under contracts ANR-11-EQPX-005-ATTOLAB and ANR-14-CE32-0011-03 APERO) and the European Research Council (under contract No. 306708, ERC Starting Grant FEMTOELEC). Access was granted to the HPC resources of CINES under allocation 2015-056057 made by GENCI. Simulations were run using EPOCH, which was developed as part of the UK EPSRC funded projects EP/G054940/1


\begin{thebibliography}{37}
\expandafter\ifx\csname natexlab\endcsname\relax\def\natexlab#1{#1}\fi
\expandafter\ifx\csname bibnamefont\endcsname\relax
  \def\bibnamefont#1{#1}\fi
\expandafter\ifx\csname bibfnamefont\endcsname\relax
  \def\bibfnamefont#1{#1}\fi
\expandafter\ifx\csname citenamefont\endcsname\relax
  \def\citenamefont#1{#1}\fi
\expandafter\ifx\csname url\endcsname\relax
  \def\url#1{\texttt{#1}}\fi
\expandafter\ifx\csname urlprefix\endcsname\relax\def\urlprefix{URL }\fi
\providecommand{\bibinfo}[2]{#2}
\providecommand{\eprint}[2][]{\url{#2}}

\bibitem[{\citenamefont{Burnett et~al.}(1977)\citenamefont{Burnett, Baldis,
  Richardson, and Enright}}]{burnett1977harmonic}
\bibinfo{author}{\bibfnamefont{N.}~\bibnamefont{Burnett}},
  \bibinfo{author}{\bibfnamefont{H.}~\bibnamefont{Baldis}},
  \bibinfo{author}{\bibfnamefont{M.}~\bibnamefont{Richardson}},
  \bibnamefont{and} \bibinfo{author}{\bibfnamefont{G.}~\bibnamefont{Enright}},
  \bibinfo{journal}{Applied Physics Letters} \textbf{\bibinfo{volume}{31}},
  \bibinfo{pages}{172} (\bibinfo{year}{1977}).

\bibitem[{\citenamefont{Carman et~al.}(1981)\citenamefont{Carman, Rhodes, and
  Benjamin}}]{carman1981observation}
\bibinfo{author}{\bibfnamefont{R.}~\bibnamefont{Carman}},
  \bibinfo{author}{\bibfnamefont{C.}~\bibnamefont{Rhodes}}, \bibnamefont{and}
  \bibinfo{author}{\bibfnamefont{R.}~\bibnamefont{Benjamin}},
  \bibinfo{journal}{Physical Review A} \textbf{\bibinfo{volume}{24}},
  \bibinfo{pages}{2649} (\bibinfo{year}{1981}).

\bibitem[{\citenamefont{Von~der Linde et~al.}(1995)\citenamefont{Von~der Linde,
  Engers, Jenke, Agostini, Grillon, Nibbering, Mysyrowicz, and
  Antonetti}}]{von1995generation}
\bibinfo{author}{\bibfnamefont{D.}~\bibnamefont{Von~der Linde}},
  \bibinfo{author}{\bibfnamefont{T.}~\bibnamefont{Engers}},
  \bibinfo{author}{\bibfnamefont{G.}~\bibnamefont{Jenke}},
  \bibinfo{author}{\bibfnamefont{P.}~\bibnamefont{Agostini}},
  \bibinfo{author}{\bibfnamefont{G.}~\bibnamefont{Grillon}},
  \bibinfo{author}{\bibfnamefont{E.}~\bibnamefont{Nibbering}},
  \bibinfo{author}{\bibfnamefont{A.}~\bibnamefont{Mysyrowicz}},
  \bibnamefont{and}
  \bibinfo{author}{\bibfnamefont{A.}~\bibnamefont{Antonetti}},
  \bibinfo{journal}{Physical Review A} \textbf{\bibinfo{volume}{52}},
  \bibinfo{pages}{R25} (\bibinfo{year}{1995}).

\bibitem[{\citenamefont{Tarasevitch et~al.}(2000)\citenamefont{Tarasevitch,
  Orisch, Von~der Linde, Balcou, Rey, Chambaret, Teubner, Kl{\"o}pfel, and
  Theobald}}]{tarasevitch2000generation}
\bibinfo{author}{\bibfnamefont{A.}~\bibnamefont{Tarasevitch}},
  \bibinfo{author}{\bibfnamefont{A.}~\bibnamefont{Orisch}},
  \bibinfo{author}{\bibfnamefont{D.}~\bibnamefont{Von~der Linde}},
  \bibinfo{author}{\bibfnamefont{P.}~\bibnamefont{Balcou}},
  \bibinfo{author}{\bibfnamefont{G.}~\bibnamefont{Rey}},
  \bibinfo{author}{\bibfnamefont{J.-P.} \bibnamefont{Chambaret}},
  \bibinfo{author}{\bibfnamefont{U.}~\bibnamefont{Teubner}},
  \bibinfo{author}{\bibfnamefont{D.}~\bibnamefont{Kl{\"o}pfel}},
  \bibnamefont{and} \bibinfo{author}{\bibfnamefont{W.}~\bibnamefont{Theobald}},
  \bibinfo{journal}{Physical Review A} \textbf{\bibinfo{volume}{62}},
  \bibinfo{pages}{023816} (\bibinfo{year}{2000}).

\bibitem[{\citenamefont{Qu{\'e}r{\'e} et~al.}(2006)\citenamefont{Qu{\'e}r{\'e},
  Thaury, Monot, Dobosz, Martin, Geindre, and Audebert}}]{quere2006coherent}
\bibinfo{author}{\bibfnamefont{F.}~\bibnamefont{Qu{\'e}r{\'e}}},
  \bibinfo{author}{\bibfnamefont{C.}~\bibnamefont{Thaury}},
  \bibinfo{author}{\bibfnamefont{P.}~\bibnamefont{Monot}},
  \bibinfo{author}{\bibfnamefont{S.}~\bibnamefont{Dobosz}},
  \bibinfo{author}{\bibfnamefont{P.}~\bibnamefont{Martin}},
  \bibinfo{author}{\bibfnamefont{J.-P.} \bibnamefont{Geindre}},
  \bibnamefont{and} \bibinfo{author}{\bibfnamefont{P.}~\bibnamefont{Audebert}},
  \bibinfo{journal}{Physical review letters} \textbf{\bibinfo{volume}{96}},
  \bibinfo{pages}{125004} (\bibinfo{year}{2006}).

\bibitem[{\citenamefont{Nomura et~al.}(2009)\citenamefont{Nomura, H{\"o}rlein,
  Tzallas, Dromey, Rykovanov, Major, Osterhoff, Karsch, Veisz, Zepf
  et~al.}}]{nomura2009attosecond}
\bibinfo{author}{\bibfnamefont{Y.}~\bibnamefont{Nomura}},
  \bibinfo{author}{\bibfnamefont{R.}~\bibnamefont{H{\"o}rlein}},
  \bibinfo{author}{\bibfnamefont{P.}~\bibnamefont{Tzallas}},
  \bibinfo{author}{\bibfnamefont{B.}~\bibnamefont{Dromey}},
  \bibinfo{author}{\bibfnamefont{S.}~\bibnamefont{Rykovanov}},
  \bibinfo{author}{\bibfnamefont{Z.}~\bibnamefont{Major}},
  \bibinfo{author}{\bibfnamefont{J.}~\bibnamefont{Osterhoff}},
  \bibinfo{author}{\bibfnamefont{S.}~\bibnamefont{Karsch}},
  \bibinfo{author}{\bibfnamefont{L.}~\bibnamefont{Veisz}},
  \bibinfo{author}{\bibfnamefont{M.}~\bibnamefont{Zepf}}, \bibnamefont{et~al.},
  \bibinfo{journal}{Nature Physics} \textbf{\bibinfo{volume}{5}},
  \bibinfo{pages}{124} (\bibinfo{year}{2009}).

\bibitem[{\citenamefont{Easter et~al.}(2010)\citenamefont{Easter, Mordovanakis,
  Hou, Thomas, Nees, Mourou, and Krushelnick}}]{easter2010high}
\bibinfo{author}{\bibfnamefont{J.~H.} \bibnamefont{Easter}},
  \bibinfo{author}{\bibfnamefont{A.~G.} \bibnamefont{Mordovanakis}},
  \bibinfo{author}{\bibfnamefont{B.}~\bibnamefont{Hou}},
  \bibinfo{author}{\bibfnamefont{A.~G.} \bibnamefont{Thomas}},
  \bibinfo{author}{\bibfnamefont{J.~A.} \bibnamefont{Nees}},
  \bibinfo{author}{\bibfnamefont{G.}~\bibnamefont{Mourou}}, \bibnamefont{and}
  \bibinfo{author}{\bibfnamefont{K.}~\bibnamefont{Krushelnick}},
  \bibinfo{journal}{Optics letters} \textbf{\bibinfo{volume}{35}},
  \bibinfo{pages}{3186} (\bibinfo{year}{2010}).

\bibitem[{\citenamefont{Borot et~al.}(2011)\citenamefont{Borot, Malvache, Chen,
  Douillet, Iaquianiello, Lefrou, Audebert, Geindre, Mourou, Qu{\'e}r{\'e}
  et~al.}}]{borot2011high}
\bibinfo{author}{\bibfnamefont{A.}~\bibnamefont{Borot}},
  \bibinfo{author}{\bibfnamefont{A.}~\bibnamefont{Malvache}},
  \bibinfo{author}{\bibfnamefont{X.}~\bibnamefont{Chen}},
  \bibinfo{author}{\bibfnamefont{D.}~\bibnamefont{Douillet}},
  \bibinfo{author}{\bibfnamefont{G.}~\bibnamefont{Iaquianiello}},
  \bibinfo{author}{\bibfnamefont{T.}~\bibnamefont{Lefrou}},
  \bibinfo{author}{\bibfnamefont{P.}~\bibnamefont{Audebert}},
  \bibinfo{author}{\bibfnamefont{J.-P.} \bibnamefont{Geindre}},
  \bibinfo{author}{\bibfnamefont{G.}~\bibnamefont{Mourou}},
  \bibinfo{author}{\bibfnamefont{F.}~\bibnamefont{Qu{\'e}r{\'e}}},
  \bibnamefont{et~al.}, \bibinfo{journal}{Optics letters}
  \textbf{\bibinfo{volume}{36}}, \bibinfo{pages}{1461} (\bibinfo{year}{2011}).

\bibitem[{\citenamefont{Hei{\ss}ler et~al.}(2012)\citenamefont{Hei{\ss}ler,
  H{\"o}rlein, Mikhailova, Waldecker, Tzallas, Buck, Schmid, Sears, Krausz,
  Veisz et~al.}}]{heissler2012few}
\bibinfo{author}{\bibfnamefont{P.}~\bibnamefont{Hei{\ss}ler}},
  \bibinfo{author}{\bibfnamefont{R.}~\bibnamefont{H{\"o}rlein}},
  \bibinfo{author}{\bibfnamefont{J.~M.} \bibnamefont{Mikhailova}},
  \bibinfo{author}{\bibfnamefont{L.}~\bibnamefont{Waldecker}},
  \bibinfo{author}{\bibfnamefont{P.}~\bibnamefont{Tzallas}},
  \bibinfo{author}{\bibfnamefont{A.}~\bibnamefont{Buck}},
  \bibinfo{author}{\bibfnamefont{K.}~\bibnamefont{Schmid}},
  \bibinfo{author}{\bibfnamefont{C.}~\bibnamefont{Sears}},
  \bibinfo{author}{\bibfnamefont{F.}~\bibnamefont{Krausz}},
  \bibinfo{author}{\bibfnamefont{L.}~\bibnamefont{Veisz}},
  \bibnamefont{et~al.}, \bibinfo{journal}{Physical review letters}
  \textbf{\bibinfo{volume}{108}}, \bibinfo{pages}{235003}
  (\bibinfo{year}{2012}).

\bibitem[{\citenamefont{Wheeler et~al.}(2012)\citenamefont{Wheeler, Borot,
  Monchoc{\'e}, Vincenti, Ricci, Malvache, Lopez-Martens, and
  Qu{\'e}r{\'e}}}]{wheeler2012attosecond}
\bibinfo{author}{\bibfnamefont{J.~A.} \bibnamefont{Wheeler}},
  \bibinfo{author}{\bibfnamefont{A.}~\bibnamefont{Borot}},
  \bibinfo{author}{\bibfnamefont{S.}~\bibnamefont{Monchoc{\'e}}},
  \bibinfo{author}{\bibfnamefont{H.}~\bibnamefont{Vincenti}},
  \bibinfo{author}{\bibfnamefont{A.}~\bibnamefont{Ricci}},
  \bibinfo{author}{\bibfnamefont{A.}~\bibnamefont{Malvache}},
  \bibinfo{author}{\bibfnamefont{R.}~\bibnamefont{Lopez-Martens}},
  \bibnamefont{and}
  \bibinfo{author}{\bibfnamefont{F.}~\bibnamefont{Qu{\'e}r{\'e}}},
  \bibinfo{journal}{Nature Photonics} \textbf{\bibinfo{volume}{6}},
  \bibinfo{pages}{829} (\bibinfo{year}{2012}).

\bibitem[{\citenamefont{Bastiani et~al.}(1997)\citenamefont{Bastiani, Rousse,
  Geindre, Audebert, Quoix, Hamoniaux, Antonetti, and
  Gauthier}}]{bastiani1997experimental}
\bibinfo{author}{\bibfnamefont{S.}~\bibnamefont{Bastiani}},
  \bibinfo{author}{\bibfnamefont{A.}~\bibnamefont{Rousse}},
  \bibinfo{author}{\bibfnamefont{J.}~\bibnamefont{Geindre}},
  \bibinfo{author}{\bibfnamefont{P.}~\bibnamefont{Audebert}},
  \bibinfo{author}{\bibfnamefont{C.}~\bibnamefont{Quoix}},
  \bibinfo{author}{\bibfnamefont{G.}~\bibnamefont{Hamoniaux}},
  \bibinfo{author}{\bibfnamefont{A.}~\bibnamefont{Antonetti}},
  \bibnamefont{and} \bibinfo{author}{\bibfnamefont{J.-C.}
  \bibnamefont{Gauthier}}, \bibinfo{journal}{Physical Review E}
  \textbf{\bibinfo{volume}{56}}, \bibinfo{pages}{7179} (\bibinfo{year}{1997}).

\bibitem[{\citenamefont{Brandl et~al.}(2009)\citenamefont{Brandl, Hidding,
  Osterholz, Hemmers, Karmakar, Pukhov, and Pretzler}}]{brandl2009directed}
\bibinfo{author}{\bibfnamefont{F.}~\bibnamefont{Brandl}},
  \bibinfo{author}{\bibfnamefont{B.}~\bibnamefont{Hidding}},
  \bibinfo{author}{\bibfnamefont{J.}~\bibnamefont{Osterholz}},
  \bibinfo{author}{\bibfnamefont{D.}~\bibnamefont{Hemmers}},
  \bibinfo{author}{\bibfnamefont{A.}~\bibnamefont{Karmakar}},
  \bibinfo{author}{\bibfnamefont{A.}~\bibnamefont{Pukhov}}, \bibnamefont{and}
  \bibinfo{author}{\bibfnamefont{G.}~\bibnamefont{Pretzler}},
  \bibinfo{journal}{Physical review letters} \textbf{\bibinfo{volume}{102}},
  \bibinfo{pages}{195001} (\bibinfo{year}{2009}).

\bibitem[{\citenamefont{Mordovanakis et~al.}(2009)\citenamefont{Mordovanakis,
  Easter, Naumova, Popov, Masson-Laborde, Hou, Sokolov, Mourou, Glazyrin,
  Rozmus et~al.}}]{mordovanakis2009quasimonoenergetic}
\bibinfo{author}{\bibfnamefont{A.~G.} \bibnamefont{Mordovanakis}},
  \bibinfo{author}{\bibfnamefont{J.}~\bibnamefont{Easter}},
  \bibinfo{author}{\bibfnamefont{N.}~\bibnamefont{Naumova}},
  \bibinfo{author}{\bibfnamefont{K.}~\bibnamefont{Popov}},
  \bibinfo{author}{\bibfnamefont{P.-E.} \bibnamefont{Masson-Laborde}},
  \bibinfo{author}{\bibfnamefont{B.}~\bibnamefont{Hou}},
  \bibinfo{author}{\bibfnamefont{I.}~\bibnamefont{Sokolov}},
  \bibinfo{author}{\bibfnamefont{G.}~\bibnamefont{Mourou}},
  \bibinfo{author}{\bibfnamefont{I.~V.} \bibnamefont{Glazyrin}},
  \bibinfo{author}{\bibfnamefont{W.}~\bibnamefont{Rozmus}},
  \bibnamefont{et~al.}, \bibinfo{journal}{Physical review letters}
  \textbf{\bibinfo{volume}{103}}, \bibinfo{pages}{235001}
  (\bibinfo{year}{2009}).

\bibitem[{\citenamefont{Thaury et~al.}(2007)\citenamefont{Thaury,
  Qu{\'e}r{\'e}, Geindre, Levy, Ceccotti, Monot, Bougeard, R{\'e}au,
  d$'$Oliveira, Audebert et~al.}}]{thaury2007plasma}
\bibinfo{author}{\bibfnamefont{C.}~\bibnamefont{Thaury}},
  \bibinfo{author}{\bibfnamefont{F.}~\bibnamefont{Qu{\'e}r{\'e}}},
  \bibinfo{author}{\bibfnamefont{J.-P.} \bibnamefont{Geindre}},
  \bibinfo{author}{\bibfnamefont{A.}~\bibnamefont{Levy}},
  \bibinfo{author}{\bibfnamefont{T.}~\bibnamefont{Ceccotti}},
  \bibinfo{author}{\bibfnamefont{P.}~\bibnamefont{Monot}},
  \bibinfo{author}{\bibfnamefont{M.}~\bibnamefont{Bougeard}},
  \bibinfo{author}{\bibfnamefont{F.}~\bibnamefont{R{\'e}au}},
  \bibinfo{author}{\bibfnamefont{P.}~\bibnamefont{d$'$Oliveira}},
  \bibinfo{author}{\bibfnamefont{P.}~\bibnamefont{Audebert}},
  \bibnamefont{et~al.}, \bibinfo{journal}{Nature Physics}
  \textbf{\bibinfo{volume}{3}}, \bibinfo{pages}{424} (\bibinfo{year}{2007}).

\bibitem[{\citenamefont{Borot et~al.}(2012)\citenamefont{Borot, Malvache, Chen,
  Jullien, Geindre, Audebert, Mourou, Qu{\'e}r{\'e}, and
  Lopez-Martens}}]{borot2012attosecond}
\bibinfo{author}{\bibfnamefont{A.}~\bibnamefont{Borot}},
  \bibinfo{author}{\bibfnamefont{A.}~\bibnamefont{Malvache}},
  \bibinfo{author}{\bibfnamefont{X.}~\bibnamefont{Chen}},
  \bibinfo{author}{\bibfnamefont{A.}~\bibnamefont{Jullien}},
  \bibinfo{author}{\bibfnamefont{J.-P.} \bibnamefont{Geindre}},
  \bibinfo{author}{\bibfnamefont{P.}~\bibnamefont{Audebert}},
  \bibinfo{author}{\bibfnamefont{G.}~\bibnamefont{Mourou}},
  \bibinfo{author}{\bibfnamefont{F.}~\bibnamefont{Qu{\'e}r{\'e}}},
  \bibnamefont{and}
  \bibinfo{author}{\bibfnamefont{R.}~\bibnamefont{Lopez-Martens}},
  \bibinfo{journal}{Nature Physics} \textbf{\bibinfo{volume}{8}},
  \bibinfo{pages}{416} (\bibinfo{year}{2012}).

\bibitem[{\citenamefont{Lichters et~al.}(1996)\citenamefont{Lichters, Meyer-ter
  Vehn, and Pukhov}}]{lichters1996short}
\bibinfo{author}{\bibfnamefont{R.}~\bibnamefont{Lichters}},
  \bibinfo{author}{\bibfnamefont{J.}~\bibnamefont{Meyer-ter Vehn}},
  \bibnamefont{and} \bibinfo{author}{\bibfnamefont{A.}~\bibnamefont{Pukhov}},
  \bibinfo{journal}{Physics of Plasmas (1994-present)}
  \textbf{\bibinfo{volume}{3}}, \bibinfo{pages}{3425} (\bibinfo{year}{1996}).

\bibitem[{\citenamefont{Liseykina et~al.}(2015)\citenamefont{Liseykina, Mulser,
  and Murakami}}]{liseykina2015collisionless}
\bibinfo{author}{\bibfnamefont{T.}~\bibnamefont{Liseykina}},
  \bibinfo{author}{\bibfnamefont{P.}~\bibnamefont{Mulser}}, \bibnamefont{and}
  \bibinfo{author}{\bibfnamefont{M.}~\bibnamefont{Murakami}},
  \bibinfo{journal}{Physics of Plasmas (1994-present)}
  \textbf{\bibinfo{volume}{22}}, \bibinfo{pages}{033302}
  (\bibinfo{year}{2015}).

\bibitem[{\citenamefont{Brunel}(1987)}]{brunel1987not}
\bibinfo{author}{\bibfnamefont{F.}~\bibnamefont{Brunel}},
  \bibinfo{journal}{Physical Review Letters} \textbf{\bibinfo{volume}{59}},
  \bibinfo{pages}{52} (\bibinfo{year}{1987}).

\bibitem[{\citenamefont{Thaury and Qu{\'e}r{\'e}}(2010)}]{thaury2010high}
\bibinfo{author}{\bibfnamefont{C.}~\bibnamefont{Thaury}} \bibnamefont{and}
  \bibinfo{author}{\bibfnamefont{F.}~\bibnamefont{Qu{\'e}r{\'e}}},
  \bibinfo{journal}{Journal of Physics B: Atomic, Molecular and Optical
  Physics} \textbf{\bibinfo{volume}{43}}, \bibinfo{pages}{213001}
  (\bibinfo{year}{2010}).

\bibitem[{\citenamefont{Kahaly et~al.}(2013)\citenamefont{Kahaly, Monchoc{\'e},
  Vincenti, Dzelzainis, Dromey, Zepf, Martin, and
  Qu{\'e}r{\'e}}}]{kahaly2013direct}
\bibinfo{author}{\bibfnamefont{S.}~\bibnamefont{Kahaly}},
  \bibinfo{author}{\bibfnamefont{S.}~\bibnamefont{Monchoc{\'e}}},
  \bibinfo{author}{\bibfnamefont{H.}~\bibnamefont{Vincenti}},
  \bibinfo{author}{\bibfnamefont{T.}~\bibnamefont{Dzelzainis}},
  \bibinfo{author}{\bibfnamefont{B.}~\bibnamefont{Dromey}},
  \bibinfo{author}{\bibfnamefont{M.}~\bibnamefont{Zepf}},
  \bibinfo{author}{\bibfnamefont{P.}~\bibnamefont{Martin}}, \bibnamefont{and}
  \bibinfo{author}{\bibfnamefont{F.}~\bibnamefont{Qu{\'e}r{\'e}}},
  \bibinfo{journal}{Physical review letters} \textbf{\bibinfo{volume}{110}},
  \bibinfo{pages}{175001} (\bibinfo{year}{2013}).

\bibitem[{\citenamefont{Bulanov et~al.}(1994)\citenamefont{Bulanov, Naumova,
  and Pegoraro}}]{bulanov1994interaction}
\bibinfo{author}{\bibfnamefont{S.~V.} \bibnamefont{Bulanov}},
  \bibinfo{author}{\bibfnamefont{N.}~\bibnamefont{Naumova}}, \bibnamefont{and}
  \bibinfo{author}{\bibfnamefont{F.}~\bibnamefont{Pegoraro}},
  \bibinfo{journal}{Physics of Plasmas (1994-present)}
  \textbf{\bibinfo{volume}{1}}, \bibinfo{pages}{745} (\bibinfo{year}{1994}).

\bibitem[{\citenamefont{Tian et~al.}(2012)\citenamefont{Tian, Liu, Wang, Wang,
  Deng, Xia, Li, Cao, Lu, Zhang et~al.}}]{tian2012electron}
\bibinfo{author}{\bibfnamefont{Y.}~\bibnamefont{Tian}},
  \bibinfo{author}{\bibfnamefont{J.}~\bibnamefont{Liu}},
  \bibinfo{author}{\bibfnamefont{W.}~\bibnamefont{Wang}},
  \bibinfo{author}{\bibfnamefont{C.}~\bibnamefont{Wang}},
  \bibinfo{author}{\bibfnamefont{A.}~\bibnamefont{Deng}},
  \bibinfo{author}{\bibfnamefont{C.}~\bibnamefont{Xia}},
  \bibinfo{author}{\bibfnamefont{W.}~\bibnamefont{Li}},
  \bibinfo{author}{\bibfnamefont{L.}~\bibnamefont{Cao}},
  \bibinfo{author}{\bibfnamefont{H.}~\bibnamefont{Lu}},
  \bibinfo{author}{\bibfnamefont{H.}~\bibnamefont{Zhang}},
  \bibnamefont{et~al.}, \bibinfo{journal}{Physical review letters}
  \textbf{\bibinfo{volume}{109}}, \bibinfo{pages}{115002}
  (\bibinfo{year}{2012}).

\bibitem[{\citenamefont{Naumova et~al.}(2004)\citenamefont{Naumova, Sokolov,
  Nees, Maksimchuk, Yanovsky, and Mourou}}]{naumova2004attosecond}
\bibinfo{author}{\bibfnamefont{N.}~\bibnamefont{Naumova}},
  \bibinfo{author}{\bibfnamefont{I.}~\bibnamefont{Sokolov}},
  \bibinfo{author}{\bibfnamefont{J.}~\bibnamefont{Nees}},
  \bibinfo{author}{\bibfnamefont{A.}~\bibnamefont{Maksimchuk}},
  \bibinfo{author}{\bibfnamefont{V.}~\bibnamefont{Yanovsky}}, \bibnamefont{and}
  \bibinfo{author}{\bibfnamefont{G.}~\bibnamefont{Mourou}},
  \bibinfo{journal}{Physical review letters} \textbf{\bibinfo{volume}{93}},
  \bibinfo{pages}{195003} (\bibinfo{year}{2004}).

\bibitem[{\citenamefont{Chen et~al.}(2006)\citenamefont{Chen, Shenga, Zheng,
  Ma, Bari, Li, and Zhang}}]{chen2006surface}
\bibinfo{author}{\bibfnamefont{M.}~\bibnamefont{Chen}},
  \bibinfo{author}{\bibfnamefont{Z.-M.} \bibnamefont{Shenga}},
  \bibinfo{author}{\bibfnamefont{J.}~\bibnamefont{Zheng}},
  \bibinfo{author}{\bibfnamefont{Y.-Y.} \bibnamefont{Ma}},
  \bibinfo{author}{\bibfnamefont{M.~A.} \bibnamefont{Bari}},
  \bibinfo{author}{\bibfnamefont{Y.-T.} \bibnamefont{Li}}, \bibnamefont{and}
  \bibinfo{author}{\bibfnamefont{J.}~\bibnamefont{Zhang}},
  \bibinfo{journal}{Optics express} \textbf{\bibinfo{volume}{14}},
  \bibinfo{pages}{3093} (\bibinfo{year}{2006}).

\bibitem[{\citenamefont{Th{\'e}venet et~al.}(2015)\citenamefont{Th{\'e}venet,
  Leblanc, Kahaly, Vincenti, Vernier, Qu{\'e}r{\'e}, and Faure}}]{thevenet2015}
\bibinfo{author}{\bibfnamefont{M.}~\bibnamefont{Th{\'e}venet}},
  \bibinfo{author}{\bibfnamefont{A.}~\bibnamefont{Leblanc}},
  \bibinfo{author}{\bibfnamefont{S.}~\bibnamefont{Kahaly}},
  \bibinfo{author}{\bibfnamefont{H.}~\bibnamefont{Vincenti}},
  \bibinfo{author}{\bibfnamefont{A.}~\bibnamefont{Vernier}},
  \bibinfo{author}{\bibfnamefont{F.}~\bibnamefont{Qu{\'e}r{\'e}}},
  \bibnamefont{and} \bibinfo{author}{\bibfnamefont{J.}~\bibnamefont{Faure}},
  \bibinfo{journal}{Nat. Phys.} p. \bibinfo{pages}{3597}
  (\bibinfo{year}{2015}).

\bibitem[{\citenamefont{Geindre et~al.}(2010)\citenamefont{Geindre,
  Marjoribanks, and Audebert}}]{geindre2010electron}
\bibinfo{author}{\bibfnamefont{J.}~\bibnamefont{Geindre}},
  \bibinfo{author}{\bibfnamefont{R.}~\bibnamefont{Marjoribanks}},
  \bibnamefont{and} \bibinfo{author}{\bibfnamefont{P.}~\bibnamefont{Audebert}},
  \bibinfo{journal}{Physical review letters} \textbf{\bibinfo{volume}{104}},
  \bibinfo{pages}{135001} (\bibinfo{year}{2010}).

\bibitem[{\citenamefont{Wang et~al.}(2010)\citenamefont{Wang, Liu, Cai, Wang,
  Liu, Xia, Deng, Xu, Leng, Li et~al.}}]{wang2010angular}
\bibinfo{author}{\bibfnamefont{W.}~\bibnamefont{Wang}},
  \bibinfo{author}{\bibfnamefont{J.}~\bibnamefont{Liu}},
  \bibinfo{author}{\bibfnamefont{Y.}~\bibnamefont{Cai}},
  \bibinfo{author}{\bibfnamefont{C.}~\bibnamefont{Wang}},
  \bibinfo{author}{\bibfnamefont{L.}~\bibnamefont{Liu}},
  \bibinfo{author}{\bibfnamefont{C.}~\bibnamefont{Xia}},
  \bibinfo{author}{\bibfnamefont{A.}~\bibnamefont{Deng}},
  \bibinfo{author}{\bibfnamefont{Y.}~\bibnamefont{Xu}},
  \bibinfo{author}{\bibfnamefont{Y.}~\bibnamefont{Leng}},
  \bibinfo{author}{\bibfnamefont{R.}~\bibnamefont{Li}}, \bibnamefont{et~al.},
  \bibinfo{journal}{Physics of Plasmas (1994-present)}
  \textbf{\bibinfo{volume}{17}}, \bibinfo{pages}{023108}
  (\bibinfo{year}{2010}).

\bibitem[{\citenamefont{Cai et~al.}(2004)\citenamefont{Cai, Gu, Zheng, Zhou,
  Yang, Jiao, Chen, Wen, and Chunyu}}]{cai2004double}
\bibinfo{author}{\bibfnamefont{D.}~\bibnamefont{Cai}},
  \bibinfo{author}{\bibfnamefont{Y.}~\bibnamefont{Gu}},
  \bibinfo{author}{\bibfnamefont{Z.}~\bibnamefont{Zheng}},
  \bibinfo{author}{\bibfnamefont{W.}~\bibnamefont{Zhou}},
  \bibinfo{author}{\bibfnamefont{X.}~\bibnamefont{Yang}},
  \bibinfo{author}{\bibfnamefont{C.}~\bibnamefont{Jiao}},
  \bibinfo{author}{\bibfnamefont{H.}~\bibnamefont{Chen}},
  \bibinfo{author}{\bibfnamefont{T.}~\bibnamefont{Wen}}, \bibnamefont{and}
  \bibinfo{author}{\bibfnamefont{S.}~\bibnamefont{Chunyu}},
  \bibinfo{journal}{Physical Review E} \textbf{\bibinfo{volume}{70}},
  \bibinfo{pages}{066410} (\bibinfo{year}{2004}).

\bibitem[{\citenamefont{Yu et~al.}(2000)\citenamefont{Yu, Bychenkov, Sentoku,
  Yu, Sheng, and Mima}}]{yu2000electron}
\bibinfo{author}{\bibfnamefont{W.}~\bibnamefont{Yu}},
  \bibinfo{author}{\bibfnamefont{V.}~\bibnamefont{Bychenkov}},
  \bibinfo{author}{\bibfnamefont{Y.}~\bibnamefont{Sentoku}},
  \bibinfo{author}{\bibfnamefont{M.}~\bibnamefont{Yu}},
  \bibinfo{author}{\bibfnamefont{Z.}~\bibnamefont{Sheng}}, \bibnamefont{and}
  \bibinfo{author}{\bibfnamefont{K.}~\bibnamefont{Mima}},
  \bibinfo{journal}{Physical review letters} \textbf{\bibinfo{volume}{85}},
  \bibinfo{pages}{570} (\bibinfo{year}{2000}).

\bibitem[{\citenamefont{Jullien et~al.}(2014)\citenamefont{Jullien, Ricci,
  B{\"o}hle, Rousseau, Grabielle, Forget, Jacqmin, Mercier, and
  Lopez-Martens}}]{jullien2014carrier}
\bibinfo{author}{\bibfnamefont{A.}~\bibnamefont{Jullien}},
  \bibinfo{author}{\bibfnamefont{A.}~\bibnamefont{Ricci}},
  \bibinfo{author}{\bibfnamefont{F.}~\bibnamefont{B{\"o}hle}},
  \bibinfo{author}{\bibfnamefont{J.-P.} \bibnamefont{Rousseau}},
  \bibinfo{author}{\bibfnamefont{S.}~\bibnamefont{Grabielle}},
  \bibinfo{author}{\bibfnamefont{N.}~\bibnamefont{Forget}},
  \bibinfo{author}{\bibfnamefont{H.}~\bibnamefont{Jacqmin}},
  \bibinfo{author}{\bibfnamefont{B.}~\bibnamefont{Mercier}}, \bibnamefont{and}
  \bibinfo{author}{\bibfnamefont{R.}~\bibnamefont{Lopez-Martens}},
  \bibinfo{journal}{Optics letters} \textbf{\bibinfo{volume}{39}},
  \bibinfo{pages}{3774} (\bibinfo{year}{2014}).

\bibitem[{\citenamefont{Borot et~al.}(2014)\citenamefont{Borot, Douillet,
  Iaquaniello, Lefrou, Audebert, Geindre, and Lopez-Martens}}]{borot2014high}
\bibinfo{author}{\bibfnamefont{A.}~\bibnamefont{Borot}},
  \bibinfo{author}{\bibfnamefont{D.}~\bibnamefont{Douillet}},
  \bibinfo{author}{\bibfnamefont{G.}~\bibnamefont{Iaquaniello}},
  \bibinfo{author}{\bibfnamefont{T.}~\bibnamefont{Lefrou}},
  \bibinfo{author}{\bibfnamefont{P.}~\bibnamefont{Audebert}},
  \bibinfo{author}{\bibfnamefont{J.-P.} \bibnamefont{Geindre}},
  \bibnamefont{and}
  \bibinfo{author}{\bibfnamefont{R.}~\bibnamefont{Lopez-Martens}},
  \bibinfo{journal}{Review of Scientific Instruments}
  \textbf{\bibinfo{volume}{85}}, \bibinfo{pages}{013104}
  (\bibinfo{year}{2014}).

\bibitem[{\citenamefont{at~[url]}()}]{Suppmat}
\bibinfo{author}{\bibfnamefont{See Supplemental Materia} \bibnamefont{at~[url]} which includes Refs \cite{jullien2009nonlinear,bocoum2015spatial,glinec2006absolute,thevenet2015}}.

\bibitem[{\citenamefont{Bocoum et~al.}(2015)\citenamefont{Bocoum, B{\"o}hle,
  Vernier, Jullien, Faure, and Lopez-Martens}}]{bocoum2015spatial}
\bibinfo{author}{\bibfnamefont{M.}~\bibnamefont{Bocoum}},
  \bibinfo{author}{\bibfnamefont{F.}~\bibnamefont{B{\"o}hle}},
  \bibinfo{author}{\bibfnamefont{A.}~\bibnamefont{Vernier}},
  \bibinfo{author}{\bibfnamefont{A.}~\bibnamefont{Jullien}},
  \bibinfo{author}{\bibfnamefont{J.}~\bibnamefont{Faure}}, \bibnamefont{and}
  \bibinfo{author}{\bibfnamefont{R.}~\bibnamefont{Lopez-Martens}},
  \bibinfo{journal}{Optics Letters} \textbf{\bibinfo{volume}{40}},
  \bibinfo{pages}{3009} (\bibinfo{year}{2015}).

\bibitem[{\citenamefont{Glinec et~al.}(2006)\citenamefont{Glinec, Faure,
  Guemnie-Tafo, Malka, Monard, Larbre, De~Waele, Marignier, and
  Mostafavi}}]{glinec2006absolute}
\bibinfo{author}{\bibfnamefont{Y.}~\bibnamefont{Glinec}},
  \bibinfo{author}{\bibfnamefont{J.}~\bibnamefont{Faure}},
  \bibinfo{author}{\bibfnamefont{A.}~\bibnamefont{Guemnie-Tafo}},
  \bibinfo{author}{\bibfnamefont{V.}~\bibnamefont{Malka}},
  \bibinfo{author}{\bibfnamefont{H.}~\bibnamefont{Monard}},
  \bibinfo{author}{\bibfnamefont{J.}~\bibnamefont{Larbre}},
  \bibinfo{author}{\bibfnamefont{V.}~\bibnamefont{De~Waele}},
  \bibinfo{author}{\bibfnamefont{J.}~\bibnamefont{Marignier}},
  \bibnamefont{and}
  \bibinfo{author}{\bibfnamefont{M.}~\bibnamefont{Mostafavi}},
  \bibinfo{journal}{Review of scientific instruments}
  \textbf{\bibinfo{volume}{77}}, \bibinfo{pages}{103301}
  (\bibinfo{year}{2006}).

\bibitem[{\citenamefont{Tarasevitch et~al.}(2007)\citenamefont{Tarasevitch,
  Lobov, W{\"u}nsche, and von~der Linde}}]{tarasevitch2007transition}
\bibinfo{author}{\bibfnamefont{A.}~\bibnamefont{Tarasevitch}},
  \bibinfo{author}{\bibfnamefont{K.}~\bibnamefont{Lobov}},
  \bibinfo{author}{\bibfnamefont{C.}~\bibnamefont{W{\"u}nsche}},
  \bibnamefont{and} \bibinfo{author}{\bibfnamefont{D.}~\bibnamefont{von~der
  Linde}}, \bibinfo{journal}{Physical review letters}
  \textbf{\bibinfo{volume}{98}}, \bibinfo{pages}{103902}
  (\bibinfo{year}{2007}).

\bibitem[{\citenamefont{Quesnel and Mora}(1998)}]{quesnel1998theory}
\bibinfo{author}{\bibfnamefont{B.}~\bibnamefont{Quesnel}} \bibnamefont{and}
  \bibinfo{author}{\bibfnamefont{P.}~\bibnamefont{Mora}},
  \bibinfo{journal}{Physical Review E} \textbf{\bibinfo{volume}{58}},
  \bibinfo{pages}{3719} (\bibinfo{year}{1998}).

\bibitem[{\citenamefont{Sheng et~al.}(2002)\citenamefont{Sheng, Mima, Sentoku,
  Jovanovi\'{c}, Taguchi, Zhang, and ter Vehn}}]{shen02}
\bibinfo{author}{\bibfnamefont{Z.-M.} \bibnamefont{Sheng}},
  \bibinfo{author}{\bibfnamefont{K.}~\bibnamefont{Mima}},
  \bibinfo{author}{\bibfnamefont{Y.}~\bibnamefont{Sentoku}},
  \bibinfo{author}{\bibfnamefont{M.~S.} \bibnamefont{Jovanovi\'{c}}},
  \bibinfo{author}{\bibfnamefont{T.}~\bibnamefont{Taguchi}},
  \bibinfo{author}{\bibfnamefont{J.}~\bibnamefont{Zhang}}, \bibnamefont{and}
  \bibinfo{author}{\bibfnamefont{J.~M.} \bibnamefont{ter Vehn}},
  \bibinfo{journal}{Phys. Rev. Lett.} \textbf{\bibinfo{volume}{88}},
  \bibinfo{pages}{055004} (\bibinfo{year}{2002}).

\bibitem[{\citenamefont{Jullien et~al.}(2009)\citenamefont{Jullien, Durfee,
  Trisorio, Canova, Rousseau, Mercier, Antonucci, Cheriaux, Albert, and
  Lopez-Martens}}]{jullien2009nonlinear}
\bibinfo{author}{\bibfnamefont{A.}~\bibnamefont{Jullien}},
  \bibinfo{author}{\bibfnamefont{C.}~\bibnamefont{Durfee}},
  \bibinfo{author}{\bibfnamefont{A.}~\bibnamefont{Trisorio}},
  \bibinfo{author}{\bibfnamefont{L.}~\bibnamefont{Canova}},
  \bibinfo{author}{\bibfnamefont{J.-P.} \bibnamefont{Rousseau}},
  \bibinfo{author}{\bibfnamefont{B.}~\bibnamefont{Mercier}},
  \bibinfo{author}{\bibfnamefont{L.}~\bibnamefont{Antonucci}},
  \bibinfo{author}{\bibfnamefont{G.}~\bibnamefont{Cheriaux}},
  \bibinfo{author}{\bibfnamefont{O.}~\bibnamefont{Albert}}, \bibnamefont{and}
  \bibinfo{author}{\bibfnamefont{R.}~\bibnamefont{Lopez-Martens}},
  \bibinfo{journal}{Applied Physics B} \textbf{\bibinfo{volume}{96}},
  \bibinfo{pages}{293} (\bibinfo{year}{2009}).

\end{thebibliography}
\end{document}